\documentclass[aps,reprint,preprintnumbers]{revtex4-1}

\usepackage{bm,latexsym,dcolumn,amsfonts,amssymb,amsmath}
\usepackage{graphicx,epsfig,footnote}
\usepackage{psfrag}
\usepackage{subfigure}
\usepackage{feynmf,rotating,hyperref}
\usepackage{color}

\hypersetup{
    bookmarks=true,         
    unicode=false,          
    pdftoolbar=true,        
    pdfmenubar=true,        
    pdffitwindow=false,     
    pdfstartview={FitH},    
    pdftitle={My title},    
    pdfauthor={Author},     
    pdfsubject={Subject},   
    pdfcreator={Creator},   
    pdfproducer={Producer}, 
    pdfkeywords={keyword1} {key2} {key3}, 
    pdfnewwindow=true,      
    colorlinks=true,       
    linkcolor=red,          
    citecolor=cyan,        
    filecolor=magenta,      
    urlcolor=green,           
    linktocpage=true
}

\newcommand{\mpl}{m_{\rm Pl}}
\newcommand{\fnl}{f_{\rm NL}}

\newcommand{\calC}{{\cal C}}

\newcommand{\calP}{{\cal P}}
\newcommand{\calR}{{\cal R}}

\begin{document}

\preprint{APCTP-Pre2015-027}

\title{Direct search for features in the primordial bispectrum}

\author{Stephen Appleby$^{a,b}$}\email{stephen@kias.re.kr}
\author{Jinn-Ouk Gong$^{b,c}$}\email{jinn-ouk.gong@apctp.org}
\author{Dhiraj Kumar Hazra$^{b,d}$}\email{dhiraj.kumar.hazra@apc.univ-paris7.fr}
\author{Arman Shafieloo$^{e,f}$}\email{shafieloo@kasi.re.kr}
\author{Spyros Sypsas$^{b,g}$}\email{ssypsas@ing.uchile.cl}
\affiliation{$^{a}$School of Physics, Korea Institute for Advanced Study, Seoul 02455, Korea}
\affiliation{$^{b}$Asia Pacific Center for Theoretical Physics, Pohang 37673, Korea}
\affiliation{$^{c}$Department of Physics, Postech, Pohang 37673, Korea}
\affiliation{$^{d}$AstroParticule et Cosmologie \& Paris Centre for Cosmological Physics, Universit\'e Paris Diderot, Paris 75205, France}
\affiliation{$^{e}$Korea Astronomy and Space Science Institute, Daejeon 34055, Korea}
\affiliation{$^{f}$University of Science and Technology, Daejeon 34113, Korea}
\affiliation{$^{g}$Departamento de F\'isica, Facultad de Ciencias F\'isicas y Matem\'aticas, Universidad de Chile, Blanco Encalada 2008, Santiago, Chile}

\begin{abstract}

We study features in the bispectrum of the primordial curvature perturbation correlated with the reconstructed primordial power spectrum from the observed cosmic microwave background temperature data. We first show how the bispectrum can be completely specified in terms of the power spectrum and its first two derivatives, valid for any configuration of interest. Then using a model-independent reconstruction of the primordial power spectrum from the Planck angular power spectrum of temperature anisotropies, we compute the bispectrum in different triangular configurations. We find that in the squeezed limit at $k \sim 0.06$ Mpc$^{-1}$ and $k \sim 0.014$ Mpc$^{-1}$ there are marginal $2\sigma$ deviations from the standard featureless bispectrum, which meanwhile is consistent with the reconstructed bispectrum in the equilateral configuration.

\end{abstract}

\maketitle

\section{Introduction}

Recent observational progress on the temperature anisotropies of the cosmic microwave background (CMB) has made primordial inflation~\cite{inflation} the most promising candidate to describe the early universe. Being an effective description at a relatively low energy embedded in an ultraviolet complete theory, it is natural to expect that the inflationary Lagrangian contains substructure which prevents otherwise smooth evolution of the universe during the entire inflationary epoch~\cite{Baumann:2014nda}. The resulting features in the primordial correlation functions imply enhanced interactions, giving rise to a unique observational window into the unknown physics of the parent theory.

Over the previous decade it has been shown that the primordial power spectrum reconstructed from the observed CMB data allows for the existence of features with roughly 5\% modulations in the amplitude~\cite{reconstruction-refs,pps-wmap,Hazra:2014jwa}. Although there is no high confidence detection beyond the smooth power-law form of the primordial power spectrum, the search for features remains tantalizing due to their potential ability to rule out a large class of inflationary scenarios. Furthermore, the existence of features in the power spectrum implies features also in higher order correlation functions. This suggests a compelling way to look for features in the bispectrum by using its correlation with the power spectrum.

In this article, using the primordial power spectrum reconstructed directly from the Planck temperature data~\cite{Hazra:2014jwa}, we search for scales and triangular configurations at which we would expect non-trivial signals in the bispectrum. The method described in this work, being both fast and accurate, provides an ideal platform to search for correlated features in the primordial bispectrum~\cite{jointsearches,Meerburg:2015owa}.

\section{Bispectrum in terms of power spectrum}
\label{sec:P-B}

The explicit correlation between the power spectrum of the curvature perturbation $\calR$ and its bispectrum was first elucidated in~\cite{Achucarro:2012fd}, where the features are sourced by a non-trivial speed of sound $c_s$. This is motivated from an effective single field description of inflation when heavy degrees of freedom are systematically integrated out~\cite{heavy}. 
Given that $\left| 1-c_s^{-2} \right| \ll 1$, we can explicitly find the leading contributions to the bispectrum $B_\calR$~\cite{Achucarro:2012fd}. Moreover, it was noted in~\cite{Gong:2014spa} that the correlation between correlation functions can be further extended in the context of the generalized slow roll formalism (GSR)~\cite{gsr} from the relation~\cite{inverse}
\begin{equation}
\log\left( \frac{1}{f^2} \right) = \int_0^\infty \frac{dk}{k} m(-k\tau) \log\calP_\calR(k) \, ,
\end{equation}
where $f = f(\log\tau) \equiv -2\pi\tau z$, with $z^2 \equiv 2a^2\epsilon\mpl^2/c_s$, $\epsilon \equiv -\dot{H}/H^2$ and $d\tau \equiv c_sdt/a$, is the GSR fundamental function and $m(x) \equiv 2\left[ x^{-1} - x^{-1}\cos(2x) - \sin(2x) \right]/\pi$. Then, the ``source'' of the power spectrum $g_P \equiv (f''-3f')/f$ with $f' \equiv df/d\log\tau$ can be written in terms of $\calP_\calR$ and its first two derivatives, i.e. $n_\calR$ and $\alpha_\calR$.

On totally general grounds, the bispectrum is specified by additional information on the source of the bispectrum $g_B$~\cite{gsrbi}, which is obtained from the cubic order action~\cite{S3}. In the effective field theory viewpoint~\cite{Cheung:2007st}, the action at each order is specified by a set of mass scales $M_n^4$, so that at quadratic order $M_2^4$ determines $c_s$, while at cubic order $M_3^4$ is to be additionally specified, in principle, as an independent coupling. This is materialized as the bispectrum source $g_B$, which is not generically written in terms of the power spectrum source $g_P$ and hence the power spectrum $\calP_\calR$ and its derivatives. Nevertheless, $g_B$ can be explicitly connected to $\calP_\calR$ if we focus our attention on the case in which the most important source of features is specified. An example is given in \cite{Achucarro:2012fd} -- a varying $c_s$ gives rise to $g_B$ that has the same origin as $g_P$, so we can specify the correlation. The same formula for $B_\calR$ can be derived using the GSR approximation~\cite{Gong:2014spa}.

Another typical and important case is when the inflaton potential $V(\phi)$ exhibits sudden changes, such as kinks and steps~\cite{footnote1,singularV,step-model}. In that case, with $c_s=1$, the dominant source of the features is the variation of $\epsilon$, i.e. $\eta \equiv {\dot\epsilon}/(H\epsilon)$ provided that inflation is not disturbed even with a violent variation of $\epsilon$ because $\epsilon \ll 1$ at all times. Then, GSR is a powerful tool to provide analytic results for the power spectrum~\cite{gsrpower} as well as the bispectrum~\cite{gsrbi}, which would otherwise require time-intensive numerical calculations~\cite{featurebi-num,bingo,Hazra:2014jwa}. The usual cubic action~\cite{S3} is, however, not appropriate because a term with $\eta$ is contained in the field redefinition, which we should later restore by hand. It is thus desirable to use an alternative form where $\eta$ is explicit outside the field redefinition terms. This form is presented in~\cite{Burrage:2011hd}, and collecting only terms with $\eta$,
\begin{equation}
\label{eq:S3}
S_3 \supset \int d^4x a^3\epsilon\mpl^2 \left[ -\eta \dot\calR^2\calR + \frac{\eta}{a^2} \calR(\nabla\calR)^2 \right] \, .
\end{equation}
To compute the bispectrum it is convenient, since $\epsilon \ll 1$, to consider a perfect de Sitter background $\tau = -1/(aH)$. Then $\eta$ can be written to leading order in GSR as
\begin{equation}
\label{eq:eta}
\eta = -\int_0^\infty \frac{dk}{k} m(-k\tau) \frac{d\log\calP_\calR}{d\log k} \, .
\end{equation}
Plugging this expression into \eqref{eq:S3} with the fluctuations given by the de Sitter mode functions, we derive an alternative form to~\cite{Achucarro:2012fd} connecting the bispectrum with the power spectrum and its first two derivatives:
\begin{widetext}
\begin{align}
\label{eq:B2}
B_\calR(k_1,k_2,k_3) & = \frac{(2\pi)^4\calP_\calR^2}{(k_1k_2k_3)^3} \left\{ \int_{K/2}^\infty dk (n_\calR-1) \frac{\Delta^2}{k^2} \right.
\nonumber\\
& \qquad\qquad\qquad + \left[ \left( k_1^2+k_2^2+k_3^2 \right) \frac{k_1k_2 + k_2k_3 + k_3k_1}{16k}  + \frac{(k_1k_2)^2 + (k_2k_3)^2 + (k_3k_1)^2}{8k} - \frac{k_1k_2k_3}{8} \right] (1-n_\calR)
\nonumber\\
& \qquad\qquad\qquad \left. + \frac{k_1k_2k_3}{8}\alpha_\calR \right\} \, ,
\end{align}
\end{widetext}
where $\calP_\calR = H^2/\left(8\pi^2\epsilon\mpl^2\right)$ is the featureless flat power spectrum and the right hand side is evaluated at $k=(k_1+k_2+k_3)/2 \equiv K/2$. Here, $\Delta^2 = K(K-2k_1)(K-2k_2)(K-2k_3)/16$ is the area squared of the triangle with three sides $k_1$, $k_2$ and $k_3$. Furthermore, with the $\fnl$ ansatz~\cite{Komatsu:2001rj} in mind, we may define a dimensionless shape function
\begin{equation}
\label{eq:fnl}
\fnl(k_1,k_2,k_3) \equiv \frac{10}{3} \frac{k_1k_2k_3}{k_1^3+k_2^3+k_3^3} \frac{(k_1k_2k_3)^2B_\calR}{(2\pi)^4\calP_\calR^2} \, .
\end{equation}
This can be evaluated in any configuration of interest. In particular, we can reproduce the standard consistency relation in the squeezed limit, say, $k_3 \ll k_1 \approx k_2$,
\begin{equation}
\label{eq:sq}
\fnl = \frac{5}{12} \left( 1-n_\calR \right),
\end{equation}
for \eqref{eq:B2}. Note that \eqref{eq:B2} is in agreement with the result of~\cite{Palma:2014hra}, where using a different method this formula was derived with the $\Delta^2$ and $k_1k_2k_3 (1-n_\calR)/8$ terms missing on the grounds that they are subleading and do not contribute in the squeezed limit.

\begin{figure*}
\includegraphics[width=0.45\textwidth]{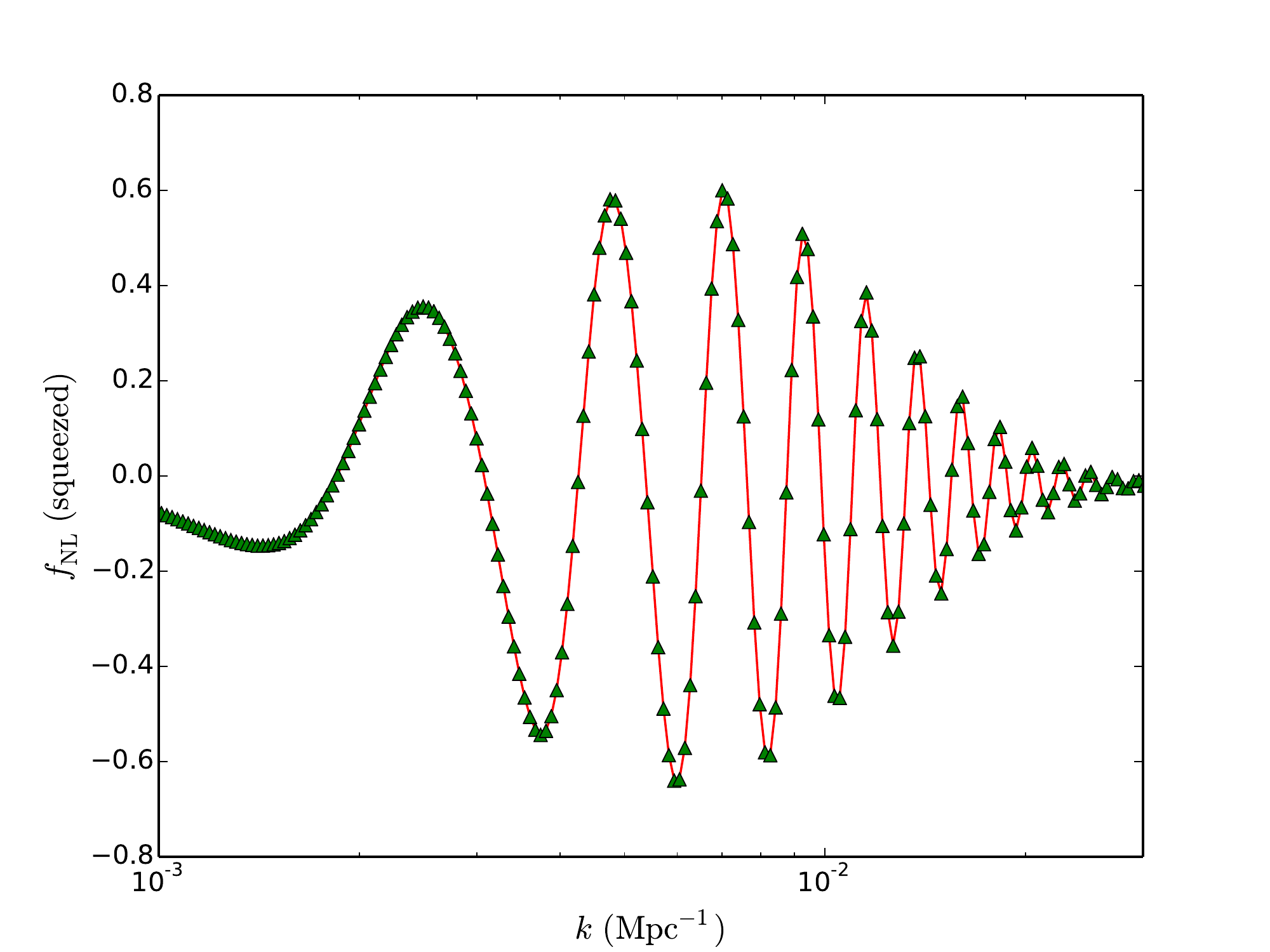}
\includegraphics[width=0.45\textwidth]{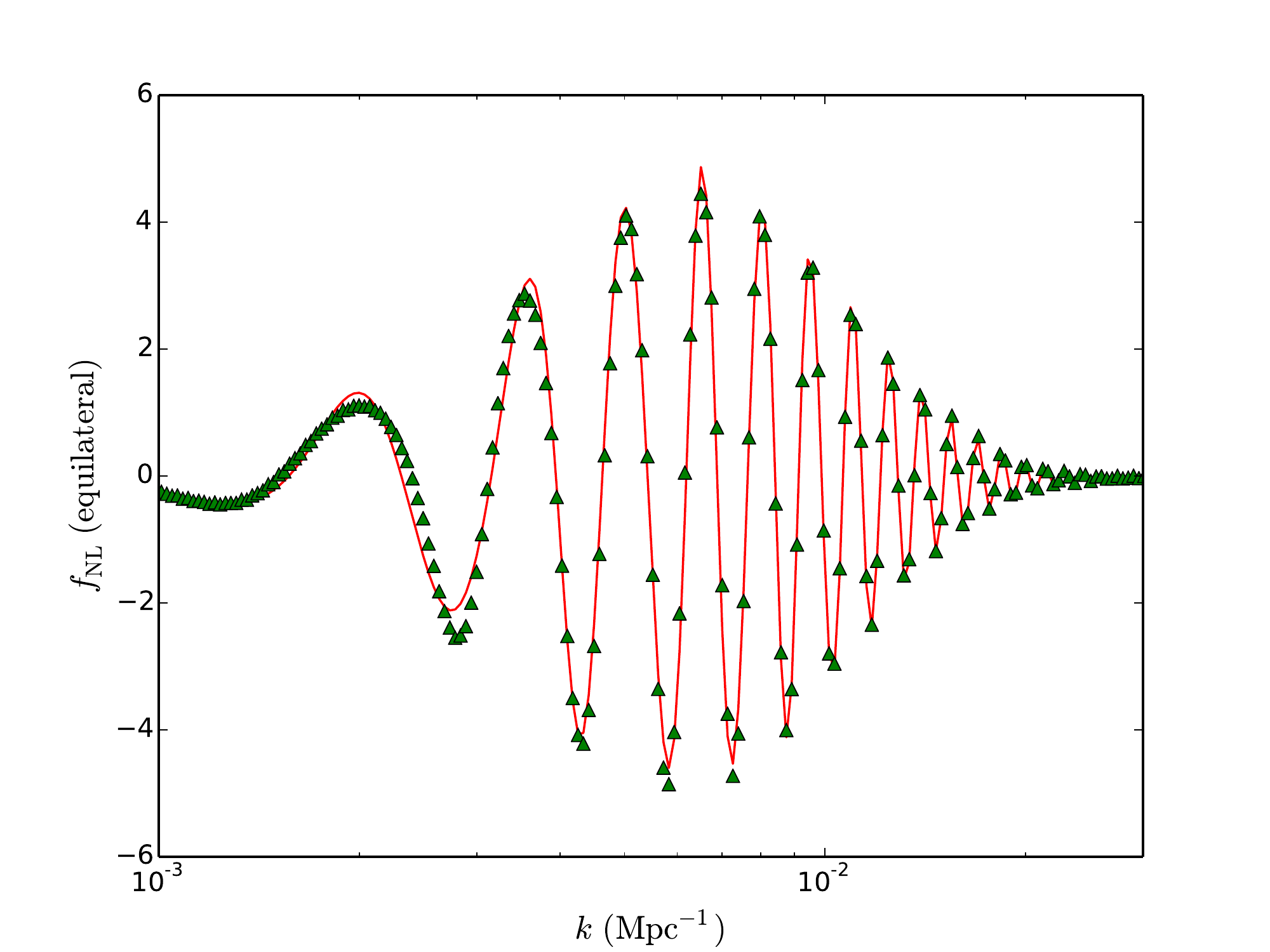}
\caption{The $f_{\rm NL}$ parameter \eqref{eq:fnl} in the (left) squeezed and (right) equilateral limit for the inflationary model \eqref{eq:stepV}. Solid lines indicate  $f_{\rm NL}$ calculated by full numerical evaluation using {\tt BINGO}, and triangles correspond to the GSR calculation \eqref{eq:fnl}.}
~\label{fig:compare}
\end{figure*}

Before applying \eqref{eq:B2} to actual data, we have tested the veracity of GSR using a step potential~\cite{step-model,featurebi-num}
\begin{equation}\label{eq:stepV}
V(\phi) = \frac{1}{2}m^2\phi^2 \left[ 1 + \alpha \tanh \left( \frac{\phi-\phi_0}{\Delta\phi} \right) \right] \, .
\end{equation}
To this end, we numerically calculated the bispectrum using {\tt BINGO}~\cite{bingo} and compared it with the output of \eqref{eq:B2}. The results are shown in Figure~\ref{fig:compare}, from which we observe excellent agreement, for parameters of the potential such that $\epsilon \ll 1$ and $\eta \lesssim 1$. Thus using GSR, if  $\calP_\calR$ can be estimated independently of any specific inflationary model, one can make general statements regarding the location and magnitude of any potentially observable features in the bispectrum. However, for sharper departures from slow-roll, where one expects slow-roll parameters to be greater than unity, our leading-order analytic approximation starts deviating from the exact solution~\cite{footnote2} and numerical computations such as {\tt BINGO} can provide the accurate results. Finally note that we can use analytic templates as well to model sharp features like steps~\cite{step-model}.

\section{Bispectrum reconstruction from power spectrum}
\label{sec:reconst}

Reconstruction of the primordial power spectrum using the CMB temperature data has been the subject of considerable research over the past decade. Here we adopt the model-independent, non-parametric modified Richardson-Lucy algorithm~\cite{pps-wmap,Hazra:2014jwa}, which relates $\calP_\calR$ in $k$-space to the CMB temperature power spectrum $\calC_\ell$ using a convolution of the form 
\begin{equation}\label{eq:RLalgorithm} 
\calC_\ell = \sum_{i} G_{{\ell}k_{i}} \calP_{k_{i}} \, , 
\end{equation}
where $G_{{\ell}k_{i}}$ is the radiative transport kernel that contains information regarding the background cosmology and $i$ denotes the discrete $k$-space binning index. $\calC_{\ell}$ is obtained using the 2013 release of Planck temperature data~\cite{Ade:2013kta}, in which we use four frequency channels covering a multipole range $2 \leq \ell \leq 2500$. The removal of foregrounds and lensing, and the iterative procedure by which~\eqref{eq:RLalgorithm} is inverted to obtain $\calP_\calR(k)$ along with all details pertaining to the reconstruction method can be found in~\cite{Hazra:2014jwa}.

The recovered $\calP_\calR(k)$ possesses both features and unavoidable noise. Indeed one cannot make a clear distinction between them when using a model-independent reconstruction. However, the important point for our purposes is that the shape of the resulting $\calP_\calR(k)$ is independent of any specific inflationary model, and the location of any significant deviations from a featureless power spectrum can direct our search for corresponding features in the bispectrum.

\begin{figure}
  \includegraphics[width=0.5\textwidth]{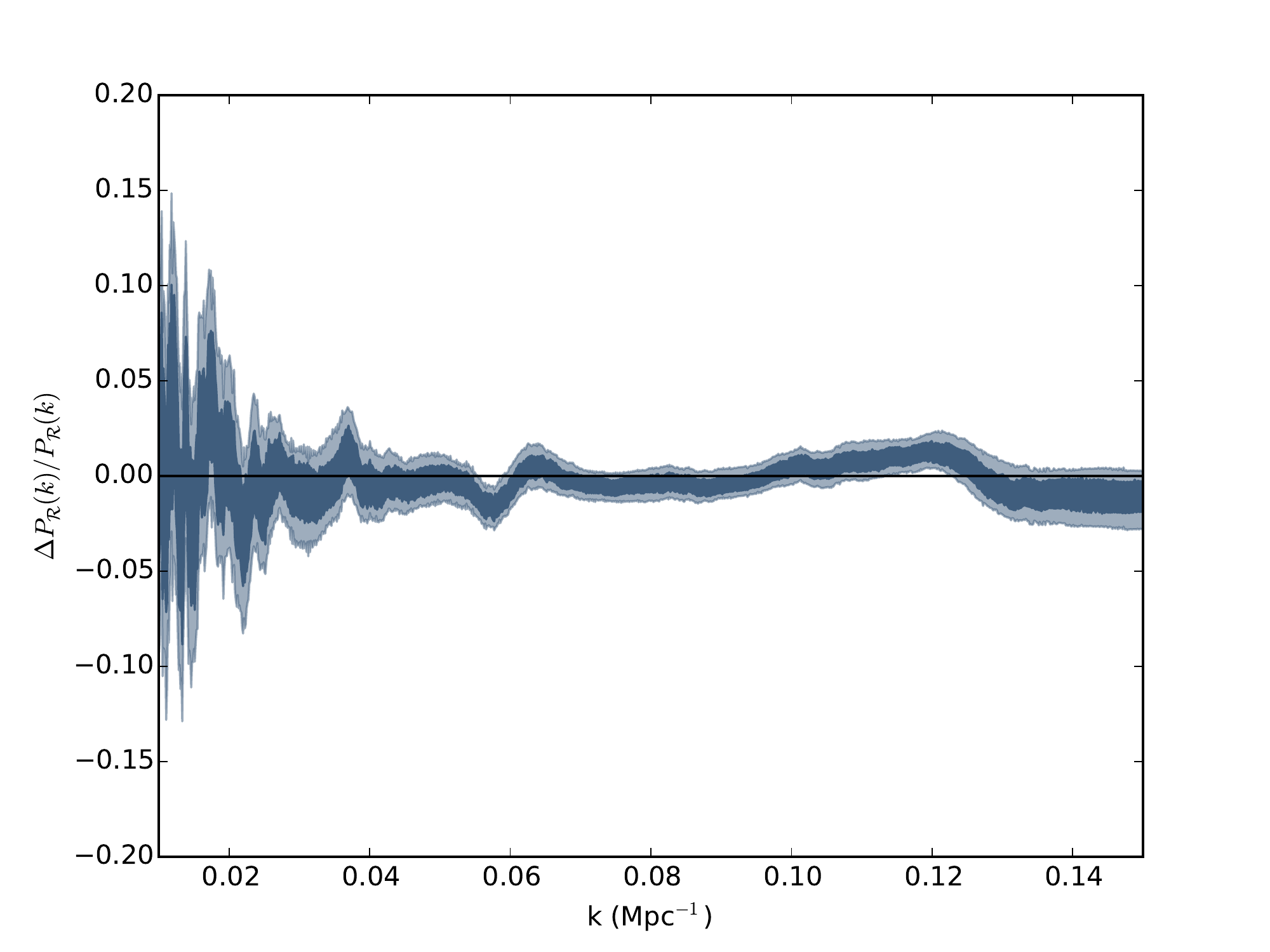}
  \caption{A model-independent reconstruction of $\calP_\calR(k)$ according to the procedure outlined in the text. Here we plot the fractional deviation with respect to the best fit power-law model. The dark (light) shaded contours contain 68\% (95\%) of the reconstructed $\calP_\calR(k)$.}
  \label{fig:spectrum}
\end{figure}

We generate $N_{\rm real}=1000$ Gaussian realizations of $\calC_{\ell}$. They are obtained by taking the existing data points and adding Gaussian random fluctuations with variance equal to the diagonal component of the full error covariance matrix. Each $\calC_\ell$ is then passed through the iterative procedure outlined in~\cite{Hazra:2014jwa} to obtain a corresponding realization of $\calP_\calR(k)$. Following this they are smoothed with a Gaussian filter, and the resulting $\calP_\calR(k)$ provides an improved fit compared to the baseline model. The result is exhibited in Figure~\ref{fig:spectrum}. As reported in~\cite{Hazra:2014jwa}, we find the maximum deviation from a featureless power spectrum lying at $k \sim 0.06~ {\rm Mpc}^{-1}$ (the other feature near $k \sim 0.12~ {\rm Mpc}^{-1}$ was reported to be a systematic error in Planck).

We now take the $N_{\rm real}=1000$ reconstructed $\calP_\calR$'s and estimate their bispectra using \eqref{eq:B2}. To do so, we must numerically calculate the first and second derivatives of $\calP_\calR(k)$. This is achieved by constructing a Chebyshev spline to approximate a smooth continuous curve using $i$ points. We use $N_{\rm Cheb}$ Chebyshev polynomials, where $N_{\rm Cheb}$ is an integer that dictates the smoothness of the splined curve. We choose $N_{\rm Cheb}$ such that the curve provides an equal $\chi^2$ fit to the Planck temperature data as the non-parametric reconstruction. Based on this criteria, we choose a value of $N_{\rm Cheb} = 1000$: a smaller value degrades the fit, resulting in a larger $\chi^{2}$ and a curve that fails to capture the important features in the data, whereas a larger value introduces spurious oscillations that amount to capturing the noise. 

\begin{figure*}[htb]
\includegraphics[width=0.4\textwidth]{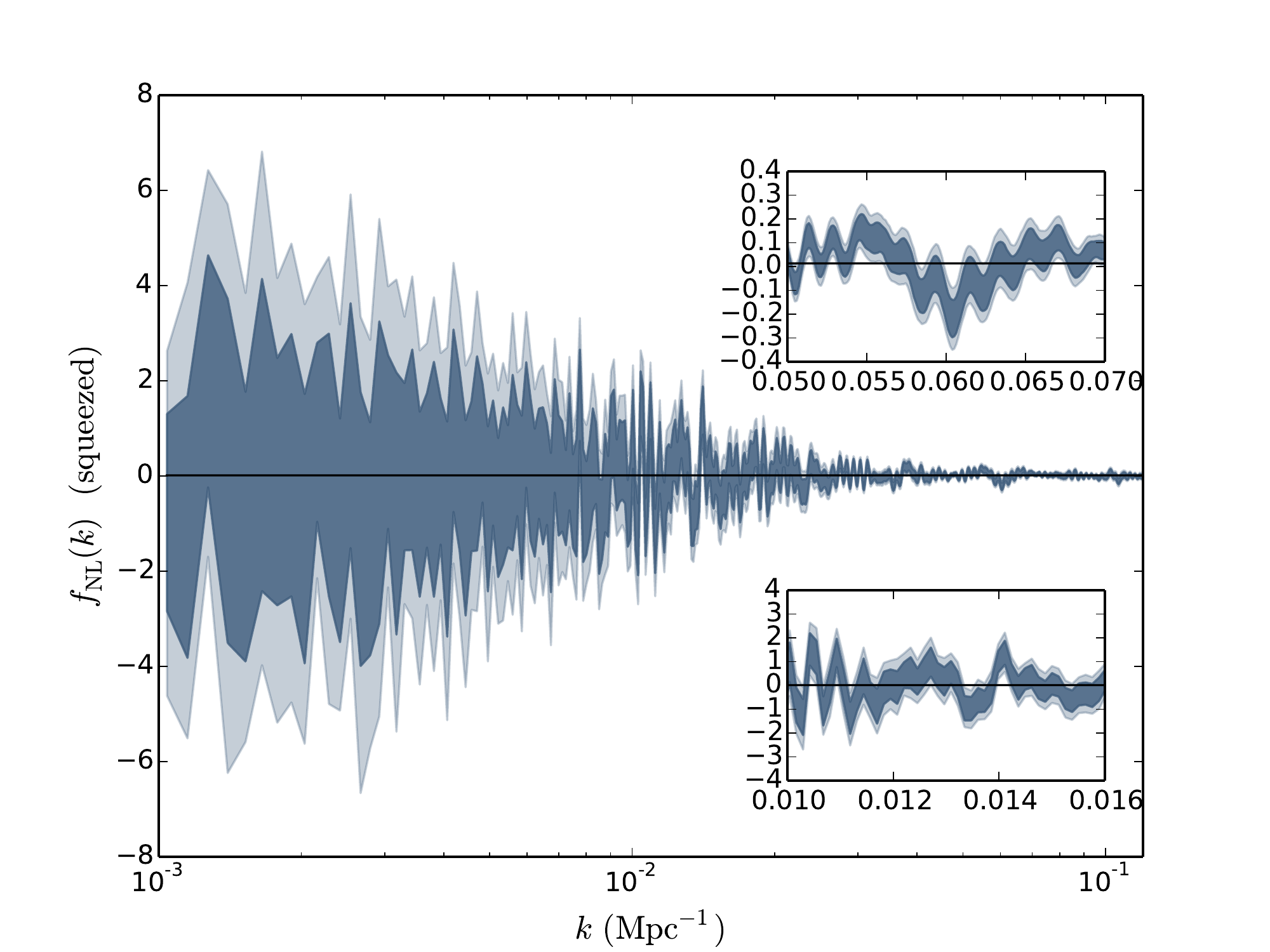}
\includegraphics[width=0.4\textwidth]{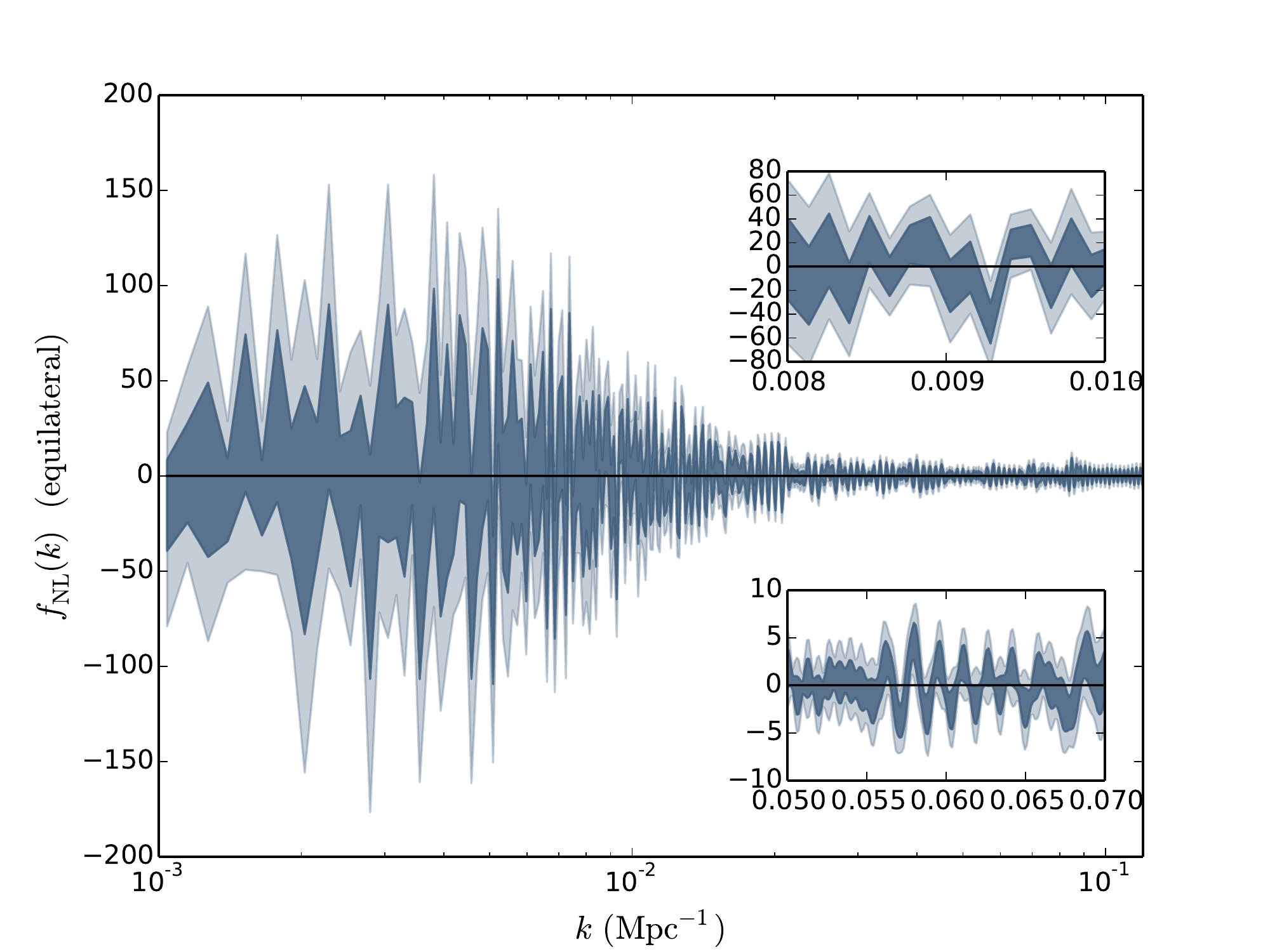}
\caption{$f_{\rm NL}$ in the (left) squeezed and (right) equilateral limit. The dark (light) band encloses 68\% (95\%) of the reconstructed $\calP_\calR$. The plot covers the entire range considered in this work, $k=(10^{-3},0.12)$ Mpc$^{-1}$. The inset plots exhibit certain $k$-bands of interest.}
 ~\label{fig:fNLproj}
\end{figure*}

Differentiating the Chebyshev fit, we calculate the bispectrum and exhibit $\fnl$ given by \eqref{eq:fnl} projected onto certain configurations in Figure~\ref{fig:fNLproj}. In the squeezed configuration, there is a mild anomalous behaviour at $k\sim 0.06~{\rm Mpc}^{-1}$, the same $k$-value for which the power spectrum exhibits a potential feature, however the amplitude of this deviation from a featureless bispectrum is very small. There is a larger amplitude discrepancy at $k \sim 0.014 ~{\rm Mpc}^{-1}$.

In the equilateral limit, the shaded region is larger than in the squeezed limit. This is expected as $\alpha_\calR = d^2\log\calP_\calR/d\log{k}^2$ is no longer suppressed for this triangular configuration. Here the bispectrum corresponding to a flat, featureless power spectrum lies within the $95\%$ bounding region for practically all $k$. There are some oscillations, however these are likely to represent noise in the fitting procedure. We arrive at this conclusion by varying $N_{\rm Cheb}$ over $N_{\rm Cheb}=(800,1200)$: all $N_{\rm Cheb}$ in this range provide a roughly comparable fit to the data and the oscillations are not robust to varying $N_{\rm Cheb}$.

\begin{figure}[!htb]
\includegraphics[width=0.45\textwidth,trim=0 0 60 0, clip]{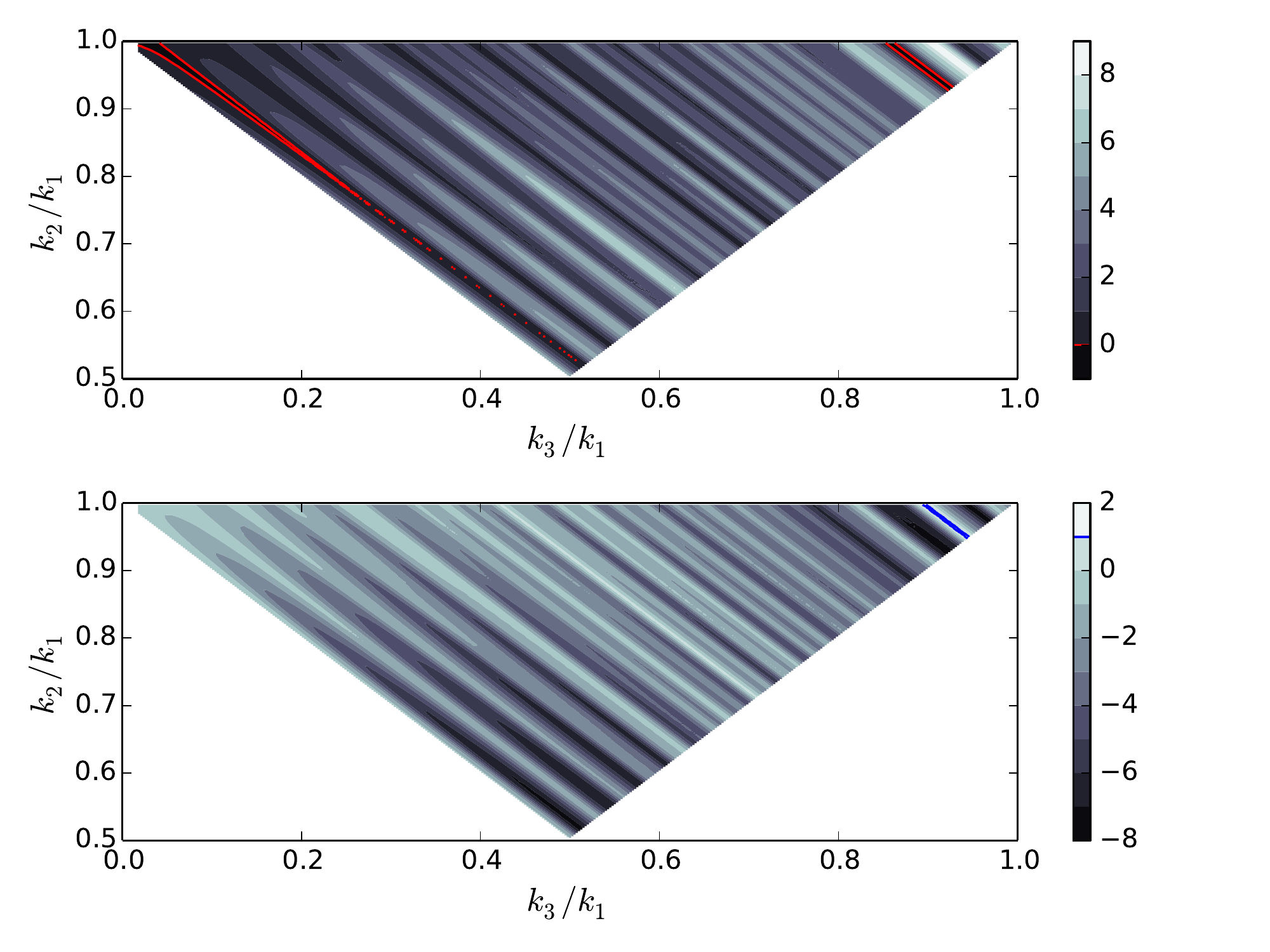}
\caption{Heat maps of $f_{\rm NL}^{+2\sigma}-f_{\rm NL}^{\rm fid}$ (top) and $f_{\rm NL}^{-2\sigma}-f_{\rm NL}^{\rm fid}$ (bottom) as a function of $k_{3}/k_{1}$ and $k_{2}/k_{1}$, with $k_{1} = 0.06 {\rm Mpc}^{-1}$. Regions of interest are $f_{\rm NL}^{+2\sigma}-f_{\rm NL}^{\rm fid} < 0$ and $f_{\rm NL}^{-2\sigma}-f_{\rm NL}^{\rm fid} > 0$, red (blue) contours in the top (bottom) panel, indicating areas where the featureless expectation value lies outside the $95\%$ contours.}
~\label{fig:fNL2D}
\end{figure}

Given that we can reconstruct $\fnl$ for any configuration, we can perform a general search in $(k_{1},k_{2},k_{3})$ space for regions with significant deviations from the flat $\calP_\calR$ expectation. We focus on $k$-bands which are known to possess a feature in $\calP_\calR$. Specifically, we fix, say, $k_{1}= 0.06 ~ {\rm Mpc}^{-1}$, and explore the residual two-dimensional $(k_{2},k_{3})$ subspace. We are especially interested in any regions where the featureless expectation value of $f_{\rm NL}$ lies outside the $95\%$ bounded region of the reconstruction. Hence for each point in the $(k_{2},k_{3})$ space, we calculate $f_{\rm NL}^{\rm fid}$, $f_{\rm NL}^{+2\sigma}$ and $f_{\rm NL}^{-2\sigma}$, where $f_{\rm NL}^{\rm fid}$ is the fiducial value of $f_{\rm NL}$, calculated for a featureless $\calP_\calR$ with $n_{\cal R} = 0.96$, and $f_{\rm NL}^{+2\sigma}, f_{\rm NL}^{-2\sigma}$ are the values that bound $95\%$ of the reconstructed $\fnl$ from above and below. In Figure~\ref{fig:fNL2D} we show $f_{\rm NL}^{+2\sigma}-f_{\rm NL}^{\rm fid}$ and $f_{\rm NL}^{-2\sigma}-f_{\rm NL}^{\rm fid}$, plotting over the range $k_{2}/k_{1}=(0.5,1.0)$ and $k_{\rm 3}/k_{1}=(10^{-3},1)$.

The stripes correspond to oscillations in $f_{\rm NL}$ observed in Figure~\ref{fig:fNLproj}. Of particular interest are regions where $f_{\rm NL}^{+2\sigma}-f_{\rm NL}^{\rm fid} < 0$ and $f_{\rm NL}^{-2\sigma}-f_{\rm NL}^{\rm fid} > 0$. These are the regions where the featureless $f_{\rm NL}$ lies outside the $95\%$ band, indicating possible features. They are exhibited as red (blue) contours in the top (bottom) panel of Figure~\ref{fig:fNL2D}. Such behaviour is most clearly observed in the top left corner of the top panel in Figure~\ref{fig:fNL2D}, i.e. most pronounced in the squeezed limit. There are a small number of narrow oscillatory bands which deviate from the featureless limit, however this is likely the same manifestation of noise as found in the equilateral configuration.

\section{Conclusion}

In this article we search for features in the bispectrum using the primordial power spectrum reconstructed from the angular power spectrum data of CMB temperature anisotropies. This is an important first step in extracting CMB three-point correlations directly from CMB two-point correlations of temperature fluctuations. The method presented is novel and allows a fast and reliable joint analysis of the CMB two- and three-point data. A model-independent reconstruction of the power spectrum can highlight the presence of features that we would expect to see in the bispectrum, and avoid model comparison with the data using a full Markov Chain Monte Carlo analysis.

Using our method, we find that in the squeezed configuration around $k \sim 0.014$ Mpc$^{-1}$ and $k \sim 0.06$ Mpc$^{-1}$ there are potential features with marginal $2\sigma$ confidence. In the equilateral configuration, the reconstructions do not strongly constrain the form of the bispectrum - a featureless bispectrum is consistent but we cannot rule out the possibility of large features being present.

Given the large uncertainty, it is important to confront these findings with CMB three-point temperature correlations directly. A joint constraint on inflationary features using the two- and three-point correlations of temperature~\cite{Meerburg:2015owa} and polarization anisotropies is the best possible approach to find or alternatively rule out features with high confidence.

\begin{acknowledgments}

We thank Jan Hamann, Gonzalo Palma and Paul Shellard for helpful discussions.
SA, JG, DKH and SS acknowledge support from the Korea Ministry of Education, Science and Technology, Gyeongsangbuk-Do and Pohang City for Independent Junior Research Groups at the Asia Pacific Center for Theoretical Physics. 
JG is also supported in part by a Starting Grant through the Basic Science Research Program of the National Research Foundation of Korea (2013R1A1A1006701) and by TJ Park Science Fellowship of POSCO TJ Park Foundation. 
DKH acknowledges Laboratoire APC-PCCP, Universit\'e Paris Diderot and Sorbonne Paris Cit\'e (DXCACHEXGS) and also the financial support of the UnivEarthS Labex program at Sorbonne Paris Cit\'e (ANR-10-LABX-0023 and ANR-11-IDEX-0005-02). 
AS would like to acknowledge the support of the National Research Foundation of Korea (2013R1A1A2013795 and 2016R1C1B2016478). 
SS is supported by the Fondecyt 2016 Post-doctoral Grant 3160299.

\end{acknowledgments}


\begin{thebibliography}{99}
\bibitem{footnote2}
See \cite{gsrbi} for complementary results, where next-to-leading order GSR corrections are shown to improve the bispectrum estimate for sharper departures from slow-roll.
\end{thebibliography}
\end{document}